\documentclass[a4paper,10pt]{article}

\usepackage[latin1]{inputenc}
\usepackage[T1]{fontenc}
\usepackage[english]{babel}
\usepackage[dvips]{graphicx} 
\usepackage{latexsym}
\usepackage{vmargin}

\setpapersize{A4}
\setmarginsrb{2.5cm}{3cm}{2.5cm}{3cm}{0cm}{1.5cm}{0cm}{1.5cm}

\title{Exo-planet detection with the COROT space mission. \\
I. A multi-transit detection criterion.}

\author{Pascal BORD\'E$^a$, Daniel ROUAN$^a$, Alain L\'EGER$^b$}

\date{}

\begin{document}

\maketitle

\begin{center}
$^a$ DESPA - CNRS, Observatoire de Paris, 92195 Meudon, France \\
Courriel : Pascal.Borde@obspm.fr, Daniel.Rouan@obspm.fr \\
$^b$ Institut d'Astrophysique Spatiale, 91405 Orsay, France \\
Courriel : Alain.Leger@ias.fr
\end{center}


\begin{abstract}
In this note, we present a detection criterion for exo-planets to be
used with the space mission COROT. This criterion is based on the transit method
that suggests to look for stars dimming caused by partial occultations by planetary
companions. When at least three transits are observed, we show that a 
cross-correlation technique can yield a detection threshold, thus enabling to
evaluate the number of possible detections assuming a model for the stellar
population in the Galaxy.
\end{abstract}

\begin{center}
\textit{Heading:} A2.~Astronomical techniques \\
\textit{Short title:} Exo-planet detection with COROT. 
\vspace*{0.5cm}
Submitted to Comptes-Rendus de l'Acad\'emie des Sciences, January 2001 / Accepted April 2001
\end{center}


\section{Overview of COROT}


\subsection{Introduction}
Exo-planets detection has now become a very active field in astrophysics.
In august 2000, about fifty Jupiter-like planets have been discovered around
nearby stars. The next challenging step is to find much less massive planets
with the hope to detect life on some of them afterwards. COROT, a CNES space
mission to be launched in 2004 [1], is currently a
funded project capable of detecting planets with a radius close to $1\,R_{\oplus}$
thanks to the ``transit method'' [2]. In this section, we briefly
describe the transit method and the COROT satellite. In section 2, we present a
detection criterion for multi-transits, and in section 3, we discuss the number
of possible dectections assuming a model for the stellar population in the Galaxy.
The specific case of mono-transits will be considered in a forthcoming article.


\subsection{The transit method}
The transit method for searching extrasolar planets is based on the idea that if
a planet crosses the disk of its parent star, it would result in a dimming of
the star's observed light [3-4]. The expected
amplitude of the relative stellar flux dimming is:
\begin{equation}
\varepsilon \equiv \frac{\Delta F}{F} = {\left(\frac{R_P}{R_{\star}}\right)}^2 
\end{equation}
where $F$ is the flux, $R_P$ the radius of the planet and $R_{\star}$ the radius
of the parent star.
We call impact parameter, expressed in stellar radii, the apparent height of the
planet trajectory above the star's equator. For an impact parameter equal to
0.5, the transit duration is:
\begin{equation}
tr = \frac{\sqrt{3}\,R_{\star}}{\sqrt{GM_{\star}}}\,\sqrt{a}
\end{equation}
where $a$ is the orbital radius. This duration is respectively 11.2 and 25.7
hours for the Earth and Jupiter in the solar system, and 3 hours for 51 Peg b
[5]. The geometrical probability that the orbital
inclination on the sky is close enough to 90\hbox{$^{\rm o}$} to make a transit
visible is:
\begin{equation}
p_g = \frac{R_{\star}}{a}
\end{equation}
$p_g$ equals to 0.5\% for the Earth, 0.1\% for Jupiter and 16\% for 51 Peg b.
This method has already been succesfully used from ground and space on the
planetary companion of HD209458, previously detected by radial velocity
techniques, e.g. [6].


\subsection{Focusing on multi-transits}

COROT features a 27 cm telescope and four $2048 \times 2048$ CCDs, two of which
are devoted to the exoplanets program. During its 2.5-year mission, the
satellite will monitor 5 fields of 5000 to 12000 stars (${11 \le m_V \le 16.5}$,
galactic latitude $=15-20^{\circ}$), each of them for 150 days, thus leading up
to 60000 lightcurves. The task will be then to look in the data for the
signatures of planetary transits. In order to estimate  the number of potential
detections by COROT, we propose to consider here a rather basic method based on
a cross-correlation technique. More accurate techniques are under study at the
Laboratoire d'Astronomie Spatial in Marseille [7], or have been
already suggested [8]. Here we will
consider only multi-transits, which means that a given planet will have to
transit at least 3 times in front of its parent star to be detected by this
method. The sought after signal looks like a repeated dimming in the parent star
lightcurve, at even time intervals corresponding to the planetary orbital period
$P$.


\section{Principle of data processing}


\subsection{Raw data averaging}
To increase the signal to noise ratio (SNR) and also to reduce the amount of
data, we begin by averaging the raw data, i.e. samples of the total flux
$F$ every 16 minutes over 150 days, on the duration of a presumed transit $tr$
(up to 15 hrs). This operation leads to a ${N=150 \times 24 / tr}$ point
vector in which a potential event is reduced to one point. It means that in the
whole process of the transit search, all relevant values of $tr$ should be tried.


\subsection{Transit signal modelling}
The transit signal $s$ we are looking for can be modeled by the sum of a Dirac comb
$\alpha\,\Pi_k$ (k teeth, amplitude $\alpha$) and a gaussian white
noise $b$ of standard deviation $\sigma_b$:
\begin{eqnarray}
\forall i \in \{1,...,N\}\,, \quad s[i] & = & -\alpha\,\Pi_k[i]+b[i] \\
\mbox{where} \quad \Pi_k[i] & = & \sum_{m=0}^{k-1}\delta[i-m\frac{N}{k}]
\end{eqnarray}
($\delta$ is Kroenecker's symbol). The term $b$ is the sum of several types
of noises:
\begin{itemize}
	\item the photon noise: $\sigma_{ph}=\sqrt{N_{ph}}$ ($N_{ph}$ refers
	actually to the number of stellar photo-electrons detected by the detector
	during $tr$)
	\item the electronical read-out noise: $\sigma_{ro}=12\,e^-.pixel^{-1}$
	\item the background noise (essentially zodiacal light):
	$\sigma_{bg}=16\,e^-.pixel^{-1}$ (exposure time is 32 s)
	\item the stellar irradiance variability noise depending on the averaging
	interval whose length is $tr$: $\sigma_{st}(tr) = N_{ph}(tr) \,
	\sigma_{st}(tr)_{ppm}$. For G and K stars, an estimated value of this quantity
	can be obtained through filtering of the SOHO-VIRGO data acquired on the Sun
	[9]: $\sigma_{st}(tr)_{ppm} \sim 30-75\,ppm$ for $tr=2-15\,hrs$.
	\item the noise introduced by random pointing error of the satellite is
	supposed to be perfectly corrected by onboard processing
\end{itemize}
For simplicity, all those noises are supposed independant, white and gaussian
noises. Consequently, we get:
\begin{equation}
\sigma_b = \sqrt{N_{ph}+n\,(\sigma_{ro}^2+\sigma_{bg}^2)+\sigma_{st}^2}
\end{equation}
where $n$ denotes the total number of read-out pixels during the transit.


\subsection{Detection with cross-correlation}
The idea is to compute cross-correlation products between the data and a Dirac
comb (amplitude unity) which has the shape of a noise free multi-transit signal
(see fig.1). For a given value of $k$, we get $N/k$ products of this kind,
denoted $C_k$:
\begin{equation}
C_k=\frac{1}{N}\sum_{i=0}^{N-1}s[i]\,\Pi_k[i]
=\frac{1}{N} \left[ \alpha\,k+\sum_{m=0}^{k-1}b[m\frac{N}{k}] \right]
\end{equation}
Trying all values of k between 3 and 50 (${3 \le P \le 50\,\mbox{days}}$) means
computing ${\sum_{k=3}^{50}N/k=\gamma N}$ products (${\gamma \simeq 3.0}$).
The presence of a transit must be somehow related to a high value of $C_k$,
but the question is: where to draw the line?


\subsection{Statistical analysis}
The answer to this question lies in a statistical analysis of the problem. To
begin with, let us consider the set of $N/k$ products $C_k$ for a given $k$.
$C_k$ can be seen as a random variable. Because we assumed that $b$ is a
gaussian noise with a null mean and a standard deviation $\sigma_b$, the
probability law of $C_k$ should also be gaussian with a standard deviation
${\sigma_C=\sqrt{k}\sigma_b/N}$. Its mean value should be equal to $\alpha k/N$
in case of a star actually showing transits or zero otherwise. Let $p_k$ be the
probability of having ${C_k \le \beta_k \, \sigma_C}$ in case of noise only (no
transit):
\begin{equation} \label{eq:pk}
p_k = Pr\{C_k \le \beta_k \sigma_C \} = \frac{1}{2}
\left[ 1+\mbox{erf} \left( \frac{\beta_k}{\sqrt{2}} \right) \right]
\end{equation} 
The probability that all the $N/k$ values of $C_k$ will remain inferior to
$\beta_k \sigma_C$ is ${a_k=p_k^{N/k}}$. This gives the level of confidence that
the statistical noise would not generate a high value of $C_k$ that would be
mistaken for a transit (see fig.2). If the statistics bear on all values of $k$
($\gamma N$ cross-correlation products), the level of confidence becomes:
\begin{equation}
a = \prod_{k=3}^{50} a_k = \prod_{k=3}^{50} {\left \{ \frac{1}{2}
\left[ 1+\mbox{erf} \left( \frac{\beta_k}{\sqrt{2}} \right) \right] \right \} }
^{N/k}
\end{equation}
One can numerically show that $\beta$ depends weakly on $k$ provided $a$ is close
enough to one (no more than 10\% variations). That is why we will assume
that $\beta$ is common to all $k$, in order to solve for $\beta$, given the
global level of confidence $a$:
\begin{equation} \label{eq:beta}
\mbox{erf} \left( \frac{\beta}{\sqrt{2}} \right) +1-2a^{\frac{1}{\gamma N}} = 0
\end{equation}
For example, the transit of a planet orbiting a sun-like star at 0.05 AU (such
as 51 Peg b) would
last approximately 3 hrs if the line of sight belongs to the orbital plane. In
this case $N=1200$, and solving (\ref{eq:beta}) for $a=99.9\,\%$ leads to
$\beta = 5.0$ (see fig.3).


\subsection{SNR detection criterion}
Assume that a global level of confidence $a$ has been chosen and has yielded up
a value of $\beta$: a detection could be claimed with that confidence, if among the
$\gamma N$ cross-correlation products, one is greater than $\beta \, \sigma_C$.
As it is a signature of a transit, its value can be written as $\alpha k / N$,
so the detection criterion translates into the inequation:
\begin{equation}
\frac{\alpha\,k}{N} \ge \frac{\sqrt{k}\,\beta\,\sigma_b}{N}
\end{equation}
Introducing the SNR on a single event (a single dimming of the lightcurve)
defined by $S/N = \alpha / \sigma_b$, we get:
\begin{equation} \label{eq:SNR}
\frac{S}{N} \ge \frac{\beta}{\sqrt{k}}
\end{equation}
As expected, the required SNR increases with the level of confidence and
decreases with the number of observed transits. To go on with the previous
example, $P=4.1\,\mbox{days}$ so $k=36$ and ${S/N \ge 0.7}$ ! As we see here, this
cross-correlation technique is enough powerful to detect repetitive transits
\textit{with lower than unity SNR on single events. }


\subsection{Link with the planetary radius}
The amplitude of the relative dimming of the parent star lightcurve is connected
to the radius of the transiting planet $R_P$ by:
\begin{eqnarray}
\frac{\Delta F}{F} & = & {\left( \frac{R_P}{R_{\star}} \right)}^2 \\
\mbox{besides} \quad \frac{\Delta F}{F} & = & -\frac{\alpha}{N_{ph}} \; = \;
-\frac{S}{N} \frac{\sigma_b}{N_{ph}}
\end{eqnarray}
Thus (\ref{eq:SNR}) translates into:
\begin{equation}
R_P \ge R_{\star} \, {\left( \frac{\beta}{\sqrt{k}} \frac{\sigma_b}{N_{ph}}
\right)}^{\frac{1}{2}}
\end{equation}
which gives the minimum planetary radius that can be detected with a level of
confidence $a$. In our example, if we assume for the star $m_V = 14$ and given
the characteristics of COROT, we compute: $R_P \ge 1.5\,R_{\oplus}$.


\section{Expected number of detections}


\subsection{Assumptions}
To have a realistic distribution of stars per spectral type and magnitude
interval in the observed fields, we have used a model of the stellar population
in the Galaxy developped at Besan\c con Observatory [10].
Only stars on the Main Sequence have been considered (type V stars). We will
assume that every observed star has 100\% probability of having a planet orbiting
at the distance tested, and that orbits are circular. For every distance, we
compute the minimum radius the planet should have for our criterion to be able to
detect it with a false alarm level of $10^{-5}$. 


\subsection{Algorithm}

Let $a$ be the orbital of the planetary orbit. We define the reduced
orbital radius by ${a_r = a {(L_{\star}/L_{\odot})}^{-0.5}}$
so that $a_r=1$ AU would always correspond to the distance where the planet
receives as much flux from its parent star as the Earth from the Sun. 
Computations have been done in the range $a_r=0.03-1$ AU.

Given the star characteristics: mass $M_{\star}$, radius $R_{\star}$, 
effective temperature $T_{\star}$, luminosity $L_{\star}$, spectral type $Sp$
and magnitude $m_V$, we compute:
\begin{itemize}
	\item the orbital radius $a$ in AU;
	\item the revolution period $P$ in days;
	\item the probability of seeing 3 transits in 150 days: $p_g \times \frac{150}{P}$
		($P \le 50$ d);
	\item the transit duration $tr$ (assuming a mean impact parameter of 0.5);
	\item the number of photo-electrons $N_{ph}$ received during $tr$.
\end{itemize}

Then, the SNR detection criterion imposes a minimum value for the signal
$\varepsilon=(S/N)/\sqrt{N_{ph}}$, and consequently for the detectable
planetary radius $R_P=R_{\star} \, \sqrt{\varepsilon}$.


\subsection{Results}
We have plotted the number of detections as a function of the reduced orbital
distance for various planetary radii (fig.\ref{fig4}) and for a false alarm rate
of $10^{-5}$. For every curve, it is the minimum detectable radius (expressed in
Earth unit) that is considered. Results are given for the whole mission. This
way of presenting our results was chosen because of the unknown frequency of the
different planetary types. Anyone can apply to our curves the scaling factor of
his choice. For instance, assuming a 2\% probability of existence, COROT could
detect several tens of ``hot Jupiters'' enhancing significantly the statistics
on this class of planets. What is more, COROT has the potential to spot ``hot
Earths'' if any, since for example around 4 events would be expected if 20\%
of the stars exhibit earth planets at 0.05 AU.


\section{References}
\begin{description}
\item{[1]} Baglin~A. et al.,
	Asteroseismology from space - The COROT experiment,
	New Eyes to See Inside the Sun and Stars, IAU 185 (1998) 301.

\item{[2]}	Rouan et al.,
	Searching for exosolar planets with the COROT space mission,
	Physics and Chemistry of the Earth Part C, v. 24, iss. 5 (1999) 567-571.

\item{[3]}	Rosenblatt~F.,
	A two-color photometric method for detection of extra-solar
	planetary systems,
	Icarus, 14 (1971) 71-93.

\item{[4]}	Schneider~J.,
	Extra-solar planets transits: detection and follow up,
	VLT Opening Symposium Antofagasta, Springer, 1999.

\item{[5]}	Mayor~M., Queloz~D.,
	A Jupiter-mass companion to a solar-type star,
	Nature, 378 (1995) 355-359.

\item{[6]}	Charbonneau et al.,
	Detection of Planetary Transits Across a Sun-like Star,
	ApJ, 529 (2000) L45-L48.

\item{[7]}	Defa\"y C. et al.,
	A bayesian method for the detection of planetary transits,
	submitted to A\&A (2000).

\item{[8]}	Jenkins et al.,
	A Matched Filter Method for Ground-Based Sub-Noise Detection of
	Terrestrial Extrasolar Planets in Eclipsing Binaries: Application to
	CM Draconis,
	Icarus, 119 (1996) 244-260.

\item{[9]}	Fr\"ohlich et al.,
	First results from VIRGO, the experiment for helioseismology
	and irradiance monitoring on SOHO,
	Solar Physics, 170 (1997) 1-25.

\item{[10]} Robin~A., Crézé~M.,
	Stellar population in the Milky Way - A synthetic model,
	A\&A, 157 (1986) 71-90.
\end{description}


\newpage

\begin{figure}[htbp]
\begin{center}

\includegraphics[height=7cm,angle=-90]{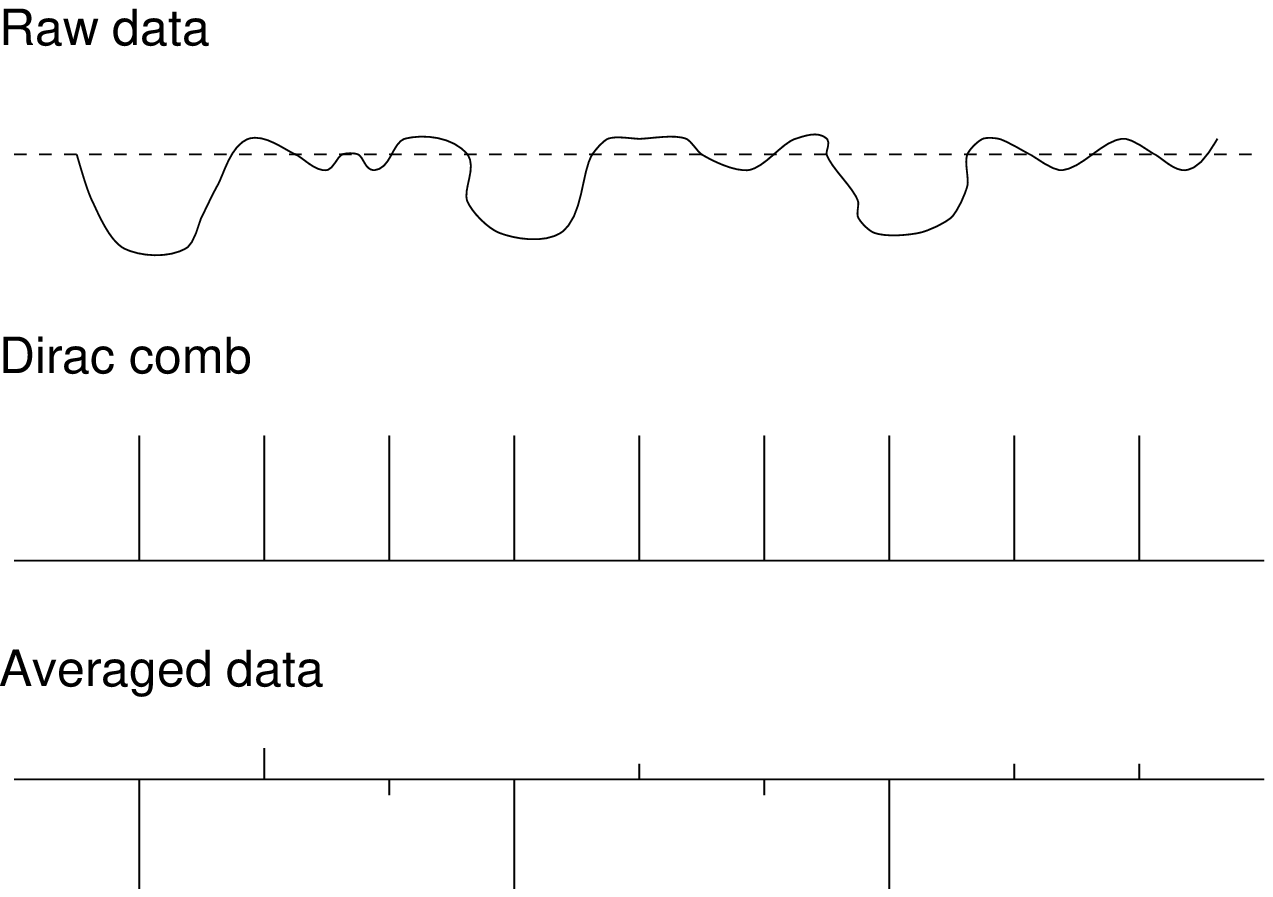}
\caption{Raw data are averaged on the duration of the presumed transit, then
cross-correlated with a Dirac Comb.}
\label{fig1}

\includegraphics[height=7cm]{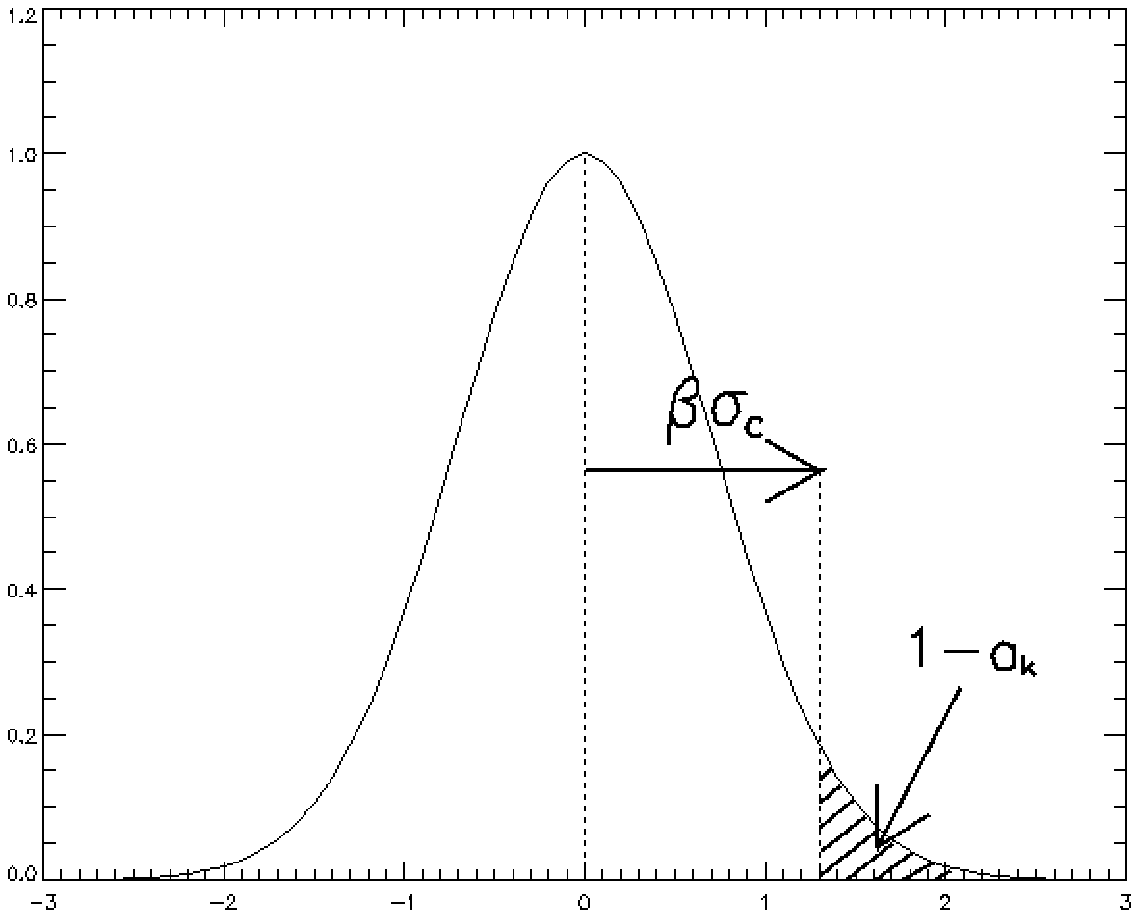}
\caption{The probability distribution of the cross-correlation products $C_k$
is centered around zero, because the absence of a transit is the general rule.
A large value of a $C_k$ with respect to $\sigma_C$ is indicative of a transit
with a level of confidence $a_k$. This confidence is also measured by $\beta$,
which is supposed common to all $k$.}
\label{fig2}

\end{center}
\end{figure}

\begin{figure}[htbp]
\begin{center}

\includegraphics[height=7cm]{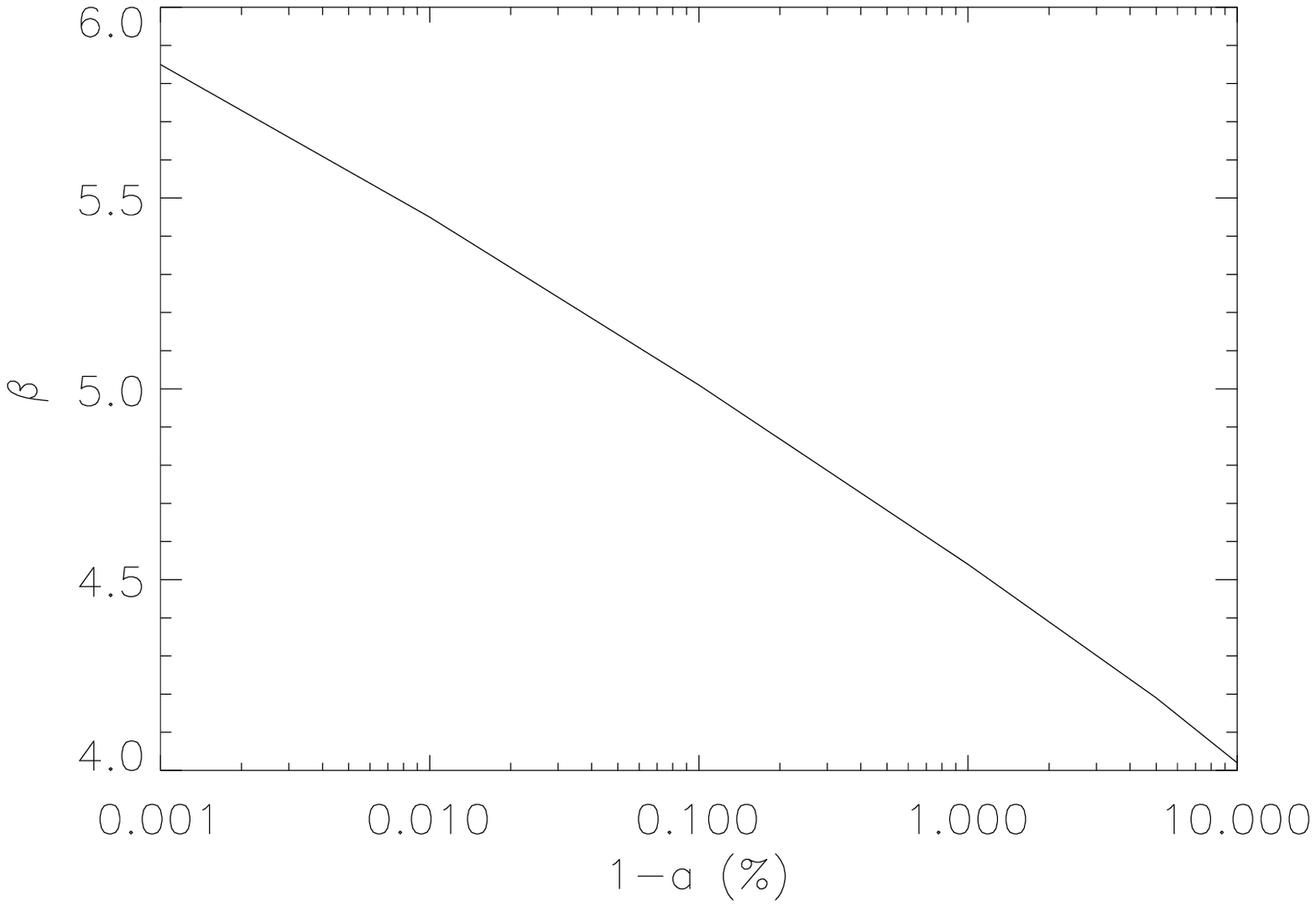}
\caption{$\beta$ as a function of the false alarm percentage $1-a$. This quasi
logarithmic relation allows to get high confidence levels with reasonable SNR.}
\label{fig3}

\includegraphics[height=7cm]{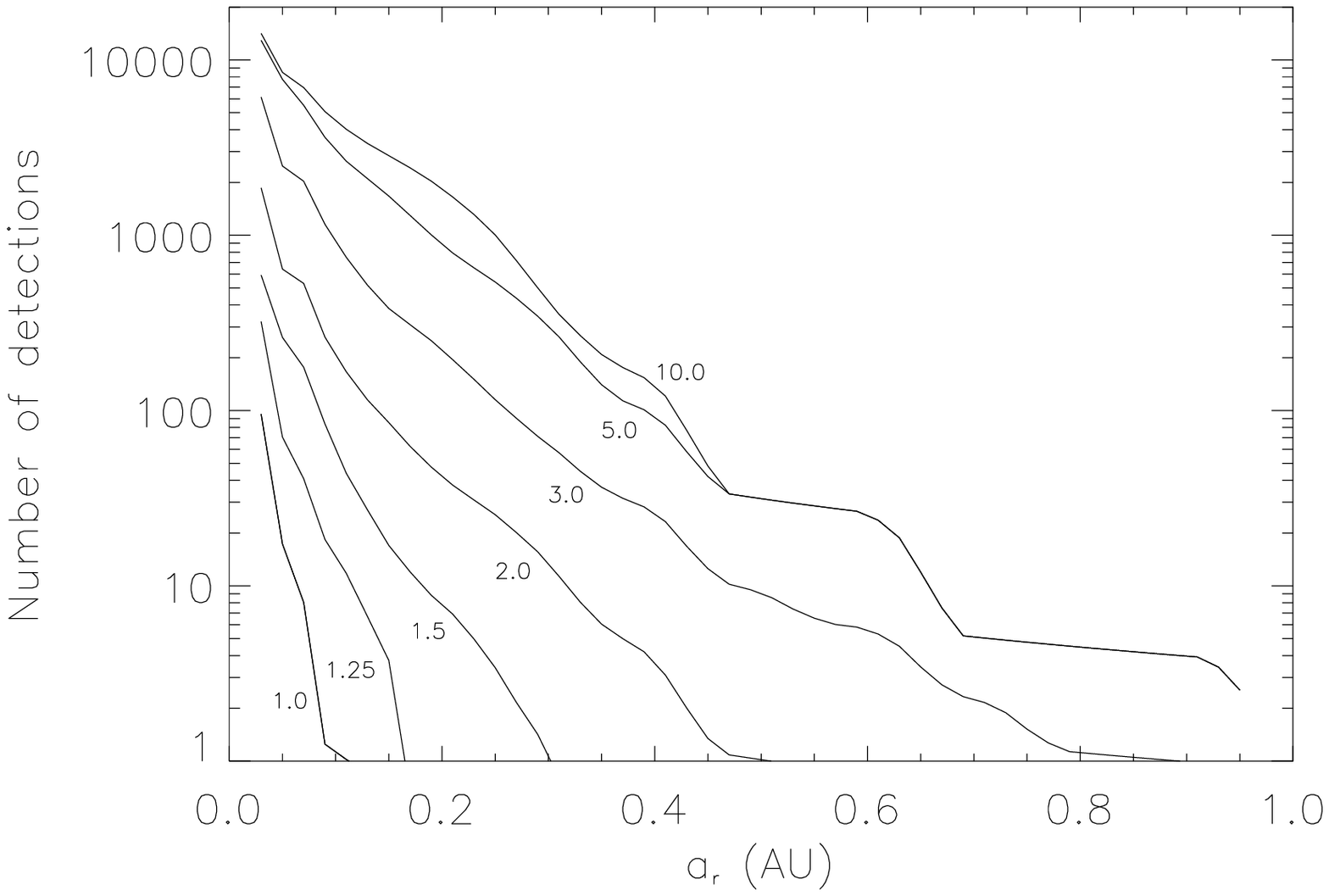}
\caption{This plot shows the number of expected detections during the full
life-time of the mission, provided 100\% of the stars have a planetary companion
of radius $R_P$ at the distance $a_r$, and assuming a false alarm rate of
$10^{-5}$. The number attached to each curve is the minimum value of $R_P$
(expressed in Earth unit) enabling the detection.}
\label{fig4}

\end{center}
\end{figure}


\end{document}